\documentclass[conference]{IEEEtran}

\usepackage{cite}
\usepackage{amsmath,amssymb,amsfonts}
\usepackage{algorithmic}
\usepackage[shortlabels]{enumitem} 
\usepackage{graphicx}
\usepackage{hyperref}
\usepackage{textcomp}
\usepackage{url}
\usepackage{xcolor}
\def\BibTeX{{\rm B\kern-.05em{\sc i\kern-.025em b}\kern-.08em
    T\kern-.1667em\lower.7ex\hbox{E}\kern-.125emX}}

\def\BibTeX{{\rm B\kern-.05em{\sc i\kern-.025em b}\kern-.08em T\kern-.1667em\lower.7ex\hbox{E}\kern-.125emX}}
\usepackage[margin=1in]{geometry}

\begin{document}

\title{HEPchain: Novel Proof-of-Useful-Work blockchain consensus for High Energy Physics\\}

\author{\IEEEauthorblockN{1\textsuperscript{st} Felix Hoffmann}
\IEEEauthorblockA{\textit{Department of Computer Science and Mathematics} \\
\textit{Goethe University}\\
Frankfurt, Germany \\
felix.hoffmann@iri.uni-frankfurt.de}
\and
\IEEEauthorblockN{2\textsuperscript{nd} Udo Kebschull}
\IEEEauthorblockA{\textit{Department of Computer Science and Mathematics} \\
\textit{Goethe University}\\
Frankfurt, Germany \\
uk@rz.uni-frankfurt.de}
}

\maketitle

\begin{abstract}
Monte Carlo simulations play a crucial role in all stages of particle collider experiments. There has been a long-term trend in HEP of both increasing collision energies and the luminosity. As a result, the requirements for MC simulations have become more rigorous: Their computational complexity has increased due to higher accuracy requirements. Additionally, more simulation data is required to allow data analysts to spot Standard Model deviations in observations of real data and enable the filtering of rare events. In order to keep up with the computational complexity of simulations and analysis of real data, distributed computing approaches are commonly employed.
For instance, CERN relies on the Worldwide LHC Computing Grid (WLCG) in order to be able to store, process, distribute and analyze collision data. Since not every HEP experiment has access to these resources and the addition of new Grid servers is a complex process, this publication explores a novel distributed computing approach for HEP which is based on blockchain technology. It features the description of a novel Proof-of-Useful-Work consensus algorithm which aims to both support real-world HEP experiments with the production of required MC data and to secure the underlying blockchain infrastructure at the same time. Instead of being an alternative to WLCG or BOINC projects that rely on volunteer computing, it aims to be a complementary source of additional computing power. This publication also features a brief introduction into blockchain fundamentals and comparisons to existing distributed computing approaches.
\end{abstract}


\section{Introduction}
Simulations in HEP are used for a variety of tasks such as R\&D of new detectors and facilities, validation and optimization of existing detectors and finally interpretation and validation of real data. Software such as FLUKA \cite{fluka_code} \cite{fluka_transport} and Geant4 \cite{geant4_recent} \cite{geant4_developments} \cite{geant4_toolkit} allow the user to define complex detector geometries and materials and are able to simulate processes such as event generation, passage of particles through detector material, digitization and reconstruction with user-adjustable levels of detail. Since full simulation takes up lots of computational power and might not be viable for complex experiments, there has been lots of work done to create hybrids of full and parameterized and machine-learning techniques to speed up bottlenecks such as calorimeter simulation which can take up a large portion of the total simulation time. At CERN, even though sophisticated fast simulation frameworks have been developed for experiments such as ATLAS or CMS, there still is a need for more computing power to be prepared for the High-Luminosity era of the LHC.
\section{The role of MC simulations in HEP}
Simulations of particle collisions play a crucial role in high-energy collider experiments. They are fundamental for:
\begin{itemize}
	\item R\&D of new detectors 
	\item Optimization of existing detectors and detector upgrades
	\item Background subtraction
	\item Interpretation and validation of real data
\end{itemize}
The previously mentioned Geant4 toolkit is a framework that allows users to simulate how particles move through detectors or rather matter in general. It has been used by large experiments at CERN such as ALICE, ATLAS, CMS or LHCb. While Geant4 in theory is capable of simulating at a high level of detail and accuracy (full simulation), in practice certain simplifications and approximations need to be made in order to speed up simulation times. Generally, there are different approaches to simulations that vary in the trade-off between computation time and accuracy. A common approach is to use full/detailed simulation (here accuracy is prioritized over computational speed) for physical processes that are especially important in the context of a given experiment  while also making use of a range of fast simulation techniques to speed up the simulation of those physical processes that are of less interest in a given context and have limited impact on results.\\\\
In the following, a condensed simulation pipeline is outlined. In the context of HEP experiments, such a process can be summarized in four steps:
\begin{itemize}
	\item (i) Event Generation
	\item (ii) Detector Response and Hit Collection
	\item (iii) Digitization
	\item (iv) Track Reconstruction
\end{itemize}
(i) In the event generation step, collisions between e.g. protons, electrons, photons or heavy nuclei at user-defined energies are simulated. In order to achieve this, known physics theory about hard and soft interaction, parton showers, decay, etc. needs to be implemented in Monte Carlo event generators such as Pythia \cite{pythia} or Dpmjet \cite{dpmjet}. These and other event generators have been added to simulation toolkits such as Geant4.\\\\
(ii) In order to simulate how the particles created in the event generation step interact with detector material, Geant4 comes with specialized modules that are able to simulate the passage of particles through different kinds of matter. There is a variety of physics lists to choose from which depending on the energy levels one is working with influence the overall computational complexity and accuracy of the simulation.\\\\
(iii) In the digitization step, energy deposits of sensitive detector material are converted into digital signals. This step is detector-specific since for some detectors the exact times of energy deposits are of interest, other detectors might have the purpose of gathering voltages or currents. Pile-up events (when a detector measures multiple events at the same time) might be too computationally expensive to be simulated in detail in the detector response step. For this reason, some digitization data might be adjusted in order to account for pile-up events. For instance, in the high luminosity era of the LHC which is expected to start in 2029, the amount of pile-up will greatly increase which will need to be handled efficiently. \cite{hl_lhc}\\\\(iv) Track reconstruction is the process of reconstructing the tracks of particles of interest by using data gathered in the previous steps. Usually a clustering approach is taken in order to aggregate detector hits that are believed to belong to the same track. Then algorithms calculate a curve that is fit to the found hits, this process is called track fitting and also provides information about the momentum of particles which combined with energy deposits measured by the detector allows for particle identification. Resulting data is reduced in multiple steps and finally written to file in ESD (Event Summary Data) or ROOT-compatible AOD (Analysis Object Data) formats which then can be analyzed by data science experts.
\subsection*{Fast simulation approaches}
Fast simulation implementations are usually detector-specific: Depending on the experiment and its goals, certain detector regions might be more important than others. In such cases a common approach is to use slow/accurate full simulations of relevant detector elements and fast/simplified geometry descriptions and simulation techniques for other detector regions that have less influence on results of interest. The following is a short list of simplifications that can be made to greatly speed up simulations:
\begin{itemize}
	\item Usage of parameterized libraries that contain pre-calculations (e.g. of frozen showers)
	\item Deep learning approaches (e.g. ATLAS FastCaloGAN)
	\item Filtering of low-energy secondaries by using an energy threshold to reduce total number of steps
	\item Choice of physics list (G4 contains various physics lists for different accuracy requirements e.g. QGSP\_BERT vs QGSP\_BERT\_HP)
\end{itemize}
It should be mentioned that the list of interactions that might affect results can be long: Electron bremsstrahlung, photon conversion, charged particle energy loss by ionization, charged particle multiple scattering, nuclear interactions, electromagnetic showers and hadronic showers are all processes that might need to be considered in the context of any HEP experiment. \cite{cms_fastsim}\\Ideally, bottlenecks of a simulation are identified and sped up using fast simulation methods. For instance, calorimeter simulation for the ATLAS experiment  at CERN used to take up around 80\% of total simulation time which led to the development of sophisticated fast simulation frameworks. The current iteration is called AtlFast3 and combines parameterized approaches with machine-learning techniques to speed-up CPU-intensive processes like calorimeter simulation. \cite{atlfast3}
\section{Distributed computing}
HL-LHC forces a revision of existing distributed computing approaches. In the following, WLCG and the LHC@home BOINC project are briefly described so that necessary future adjustments can be identified. Additionally, a novel approach called HEPchain is presented which can be a promising complementary source of computing power in the future.

\subsection{WLCG}
The Worldwide LHC Computing Grid (WLCG) is a collaboration of approximately 170 computing centers in more than 40 countries which was established in 2001. While the main goal of the grid is to store, process, distribute and analyze collision data created at ALICE, ATLAS, CMS and LHCb, many other experiments are also able to use Grid resources. Combined there are a total of around 1.4 million cores and more than 1500 petabytes of storage available, CERN itself provides approximately a fifth of the available resources. \cite{cern_grid}\\\\
Grid data centers are divided into a hierarchical system:
\begin{itemize}
	\item Tier 0: High-performance computing (HPC) systems owned and maintained by CERN. Responsible for real-time processing and reduction of raw data produced by LHC experiments.
	\item Tier 1: Large data centers (with regard to available cores and storage) connected to tier 0 data centers via dedicated fiber links (10-100 Gbit/s). Responsible for initial processing, analysis and long-term archiving of data.
	\item Tier 2: Intermediate-scale data centers that receive data from tier 1 centers and are responsible for further processing and analysis of data. These commonly are owned, maintained and used by individual university groups and scientific institutions.
	\item Tier 3: Smaller computing centers that can receive data from tier 2 centers. Just like tier 2 centers, they are commonly used by researchers for data analysis.
\end{itemize}
Without the WLCG, CERN would not be able to handle the large amounts of data produced by the LHC.  Maintaining such a complex distributed computing grid also brings challenges that need to be overcome:
\begin{itemize}
	\item Setting up new grid sites is a complex process
	\item Cost of maintaining hardware (frequent upgrades required)
	\item Reliance on high-bandwidth networks. Network issues need to be monitored in real-time to prevent delays in data processing and analysis.
	\item Secure transmission of large amounts of potentially sensitive data.
	\item Fault-tolerant data management is required to prevent loss of data. Determining a suitable degree of redundancy is non-trivial in complex, distributed systems such as the WLCG.
\end{itemize}

\subsection{BOINC approach}
BOINC (Berkeley Open Infrastructure for Network Computing) is a platform for volunteer computing which allows individuals to donate available computational power to support real-world experiments. After downloading the official BOINC client \cite{boinc_client}, the user can choose from a list of scientific experiments to support. After choosing a project to support, there might be a required download e.g. in the form of a VM that contains all required software and dependencies. Finally, a so-called \textit{work unit} is downloaded which is a computing task that is to be solved by the user. Typically, a difficult problem is split up into many work units of varying difficulty. This is done to make efficient use of volunteer computing environments that hardware-wise can be described as highly heterogeneous networks. Once the work is completed, the results are sent back to a server that collects and verifies results from all volunteer supporters of a given experiment. In the following, the default verification approach is described.

\subsubsection*{Verification of results}
Even though every project can use custom verification mechanisms, BOINC by default relies on a so-called \textit{verification-by-replication} approach. \cite{boinc_redundancy} \cite{boinc_validation} The basic idea of this approach is to hand out the same problem to many random users and then assume that the result that has been submitted most often is the correct solution.\\
There is a variable called \textit{min\_quorum} which is equal to the amount of users that must submit the same result until this result is considered to be the correct solution. Additionally, there is a variable called \textit{target\_nresults} which is the amount of correct solutions that need to be gathered until this work-unit is considered to be completed, at which point is will not be sent out to additional users. Naturally it follows that $target\_nresults \geq min\_quorum$.
Since floating-point operations can lead to slightly different results across users due to different hardware and possibly varying math libraries, it is possible to add tolerances so that any solution near a set threshold can be viewed as being identical. \cite{boinc_paper}\\

\subsubsection*{LHC@home}
The LHC@home BOINC project \cite{lhc_athome} started 2004 and allows users to run simulations to improve the design of LHC and its detectors. As of January 2023, it's average compute power is estimated to be around 31 TeraFLOPS. \cite{lhc_athome_stats}\\\\
Within this project users are, as of December 2022, able to choose between the following applications:
\begin{itemize}
	\item ATLAS (long simulation)
	\item ATLAS Simulation
	\item CMS Simulation
	\item SixTrack
	\item Sixtracktest
	\item Theory Simulation
\end{itemize}
While the ATLAS and CMS applications run detector-specific simulations to support these experiments, SixTrack focuses on improving the stability of the beam itself. More specifically, it simulates the proton stability of 60 particles that travel through LHC for more than 100000 loops which corresponds to a time frame of around ten seconds in the real world. \cite{sixtrack}\\
Theory Simulation (which used to be called Test4Theory) in 2011 became the world's first VM-based BOINC project and reached four trillion simulated events in 2018. \cite{test4theory_milestone} Its goal is the theoretical fitting of past experimental data to Standard Model expectations. This is done by comparing real LHC data to MC simulations that make use of theoretical models that are based on known theory. Results are added to a public database that is part of the MCPlots project which is an online repository of MC plots of event generators. Its goal is to provide visualized comparisons of simulated data to a wide range of real experimental data. \cite{mcplots}\\
In the past there has also been an application called Beauty@LHC for the LHCb experiment \cite{lhcb_beauty}), but it currently is suspended. Even though there have been theoretical proposals such as ALICE Connex \cite{alice_connex}, ALICE remains the only large CERN experiment that never had a realized BOINC project.\\\\
All in all, LHC@home not only raised public awareness of experiments at CERN but also attracted volunteers from all around the world to support the experiments by donating computing power. In the context of CERN, resuming the currently suspended LHCb Beauty project and creating a new LHC@home sub-project for ALICE are measures that can be taken to provide additional computing power when it is needed in the future.

\section{Blockchain consensus fundamentals}
In the following, blockchain consensus algorithm fundamentals and properties of hash-based PoW are described. It is also outlined why these properties are useful in the context of a blockchain. 

\subsection{Introduction to Proof-of-Work (PoW)}
\textit{Proof-of-Work} (PoW) is a blockchain consensus algorithm in which miners repeat hash operations with certain inputs in order to solve the hash puzzle of the current block. PoW not allows nodes that do not trust each other to find consensus but also secures the integrity of the underlying blockchain. The most popular (as in total hash rate of the network) PoW blockchain currently is the cryptocurrency Bitcoin which often is criticized for its waste of energy due to PoW consensus. The energy used for repeating hash operations has no positive benefit outside of the blockchain. Thus, in recent years a modified consensus algorithm called \textit{Proof-of-Useful-Work} has been the focus of research. The idea here is that the hash puzzle from PoW is replaced by calculations that support real-world scientific experiments. Not every problem is suitable to be used as the basis of such a consensus algorithm, since certain properties that the hash-based PoW brings need to be retained. The next section lists these useful properties and outlines how they can retained in PoUW.

\subsection{Properties of PoW}
\begin{enumerate}
	\item Block sensitivity
	\item Adjustable problem hardness
	\item Parallelization
	\item Fast verification
\end{enumerate}
1) Block sensitivity means that the problems that need to be solved by the miners to create a new block must be bound to a certain period in time. This serves two purposes: Future solutions can't be pre-calculated and previous solutions can't be re-used.\\\\
2) Adjustable problem hardness is necessary so that the difficulty of the problems dynamically can be adjusted depending on the current hash rate of the network. If the underlying blockchain also features a cryptocurrency like in Bitcoin, then this property is required to facilitate a stable transaction throughput.\\\\
3) Parallelization enables the formation of mining pools in which multiple nodes work together on the problem and share block rewards proportionally to the individual node hash rates. This is a useful property because it allows entities with old hardware to participate in the blockchain without having to expect a zero income with a high probability. This property lowers entry barriers which as a result helps in securing the blockchain due to a higher average network hash rate. The higher the hash rate of the network, the more expensive it is to attack it using a so-called 51\% attack in which a single entity controls more than half of the network hash rate which would allow it re-write the history of the blockchain.\\\\
4) Fast verification is required so that nodes do not have to rely on trust but instead can efficiently verify the proposed solutions by other miners. The problem that is to be solved usually is a one-way function such as e.g. a cryptographic hash function like SHA256. This way the problem itself can be difficult but given a solution, its validity can quickly be determined.

\section{HEPchain}
In this section, a novel approach for distributed computing in HEP called HEPchain is presented and its advantages and potential disadvantages compared to traditional distributed computing approaches are analyzed. In contrast to BOINC projects which are purely volunteer without monetary rewards, blockchain-based approaches can be coupled with a cryptocurrency that can give a financial incentive for miners to provide their computational resources. It is explained which useful properties of hash-based PoW can be retained by HEPchain and what potential challenges for its implementation will be. The process of setting up a node should be simplified by aggregating all required simulation software and dependencies in a Virtual Machine that can be downloaded. This way, little technical expertise is required which lowers the entry barriers of the blockchain network.
\subsection{Overview}
HEPchain is a blockchain construct that uses a combination of a Proof-of-Useful-Work consensus algorithm and a root authority approach to reward useful work, perform transaction and secure the underlying blockchain. Let's assume real-world HEP projects A, B and C decide to form a root authority and combine some of their computational resources to form root nodes. Root nodes are responsible for collecting solutions from miners, verifying them and for rewarding correct solutions in the form a HEPtoken which is the underlying currency of the blockchain. The state of the blockchain is synced between all nodes and contains all blockchain data except for the simulation data itself (to reduce the size of the blockchain). The following information is stored in every block of the chain:
\begin{itemize}
	\item Block number
	\item Block creation timestamp
	\item Hash of the previous block
	\item List of transactions
	\item Address of the node that received block reward
	\item List of MC simulation parameters
	\item Hash of the simulation data that this node provided
\end{itemize}
Using the block structure described above, the size of the blockchain itself will be limited in size which allows new nodes to enter the network and complete a full sync which results in a high degree of decentralization.

\subsubsection{Storage of simulation data}
Projects A, B and C run a centralized server that stores all verified solutions of miners in {Hash: Data} tuples. This way nodes can request simulation data given the hash if desired, but this feature can also be disabled by A, B and C to save bandwidth or for request spam protection. Additionally, the root authority runs a database that has direct access to the current token balance of all nodes. This is done by executing every transaction including in a block when it is published. Nodes can send requests to the root authority to be informed about the current token amount of a given address. Alternatively, a node can manually verify this information by starting at the genesis block (Block 0) and manually repeating all transactions of each block to reach the current state. This allows the node to run lookup operations locally instead of being forced to trust information given by the root authority.

\subsubsection{Block creation and rewards}
In order to provide a Proof-of-Useful-Work, miners run MC simulations with pre-defined parameters. The root authority defines a new set of simulation parameters for every block and broadcasts them to the network. Any submitted solution that uses a different set of parameters is invalid. Block creation is supposed to happen in regular intervals (this helps to stabilize transaction throughput and to prevent token inflation from block rewards), but these intervals can be adjusted in length if projects A, B and C require a higher/lower degree of accuracy in simulations. The amount of time available to solve the current problem should depend on the estimated problem difficulty and the amount of miners currently available in the blockchain network. After there is no more time left to solve the current problem, the root authority validates every solution submitted by the miners. A winner (the miner that receives the block reward token) is chosen randomly from all valid solutions, then a root authority creates a new block and broadcasts it to the network. Included in this block also is a list of all transactions that were broadcast by nodes. Since the root authority profits from the simulation data itself, there are no transaction costs necessary. The mechanism of paying higher transaction fees to have transactions conducted faster like it is found in existing blockchain-based cryptocurrencies does not exist and is not required, because block intervals are regular and the size of the transaction list is not limited. In case spamming low value transactions becomes a problem due to bloating the size of the blockchain, a transaction list size cap can be introduced which would delay all incoming transactions to the next block in case the current block is already filled.\\
Since the address of the root authority is known, all nodes will acknowledge new blocks broadcast by it and sync it to their locally stored blockchain. In order to prevent Sybil attacks (in which a miner submits the same solution from many different identities to increase the probability of being selected as a winner) nodes can only become miners by verifying their real-world identity with the root authority. Every miner can only submit one solution per block, the simulation data itself does not have to be encrypted because it is only sent directly to the root authority servers whose addresses are known. Therefore, miners are not able to copy the solutions of other miners and upload them.

\subsubsection{Prevention of spam}
Submitting incorrect solutions (faulty calculations or wrong parameter list) can lead to a ban from participating in the blockchain. This is possible because the real identities of miners and their sender addresses are known by the root authorities.

\subsubsection{Validation of results}
Since the root authority needs to be able to verify the correctness of proposed problem solutions efficiently, there is a need for a (probabilistic) verification algorithm. In this publication, multiple verification approaches are described and compared with each other:
\begin{enumerate}[a)]
	\item Verification-by-replication
	\item Pre-calculation of sub-problems and Clustering
	\item Comparison with real data
\end{enumerate}
The first approach is inspired by the default BOINC verification mechanism: After collecting all proposed solutions for a given block, the root authority will determine the most common solution and assume that since many different miners came to this result is should be the correct solution. An advantage of this approach is that it is easy to implement and that little computational effort is required. The downside of this approach is that there is no guarantee whether the accepted solution is correct and it opens up Sybil attack vectors in which either one entity acquires and registers many identities and uses them to submit identical solutions or in which multiple entities collaborate together to publish the same fake solution.\\\\
The second approach is based so-called decoys which are a parts of the problem whose solutions are already known by the root authority. If, for instance, the current block problem consists of running the simulation pipeline (Event Generation, Detector Response and Hit Collection, Digitization and Track Reconstruction) with a given RNG seed for many different configuration flags, then the root authority can randomly choose one of these configurations and run the simulations themselves. Now this part of every proposed solution by miners can be compared to this reference solution to verify its validity. If this part of the solution is correct, it can be assumed that likely all other configurations are also correct. Since now it is possible to generate a set of probably correct solutions, the most common solution among themselves can now be selected as correct solution (with high certainty). Clustering approaches can be used to determine which solutions are most similar to a given one. Finally, select one random miner from the list of all miners that uploaded this exact solution and reward a block token. The advantage of this approach is that generating fake solutions becomes more difficult since they are easily spotted if they are completely false. The only thinkable attack would be that many different miners collaborate to publish identical solutions that are made-up partly and partly correct. This would give them a non-zero chance to have correctly simulated the part of the problem that the root authority will use for validation. Many identities would be required so that the most common solution would be the one the group of malicious miners have conspired on. The disadvantage of this approach is that the root authorities need to have the computational resources to simulate part of the solution themselves. The seed required for the simulations is not known beforehand because it depends on the hash of the previous block, which is why these decoys can not be pre-calculated ahead of time. Additionally, while this approaches greatly decreases the probability of selecting a solution the is not entirely correct as the winner, it is not infallible.\\\\
The third approach is to compare simulation data provided by the nodes to real data that has been collected from detectors of a HEP experiment. Kalman filtering can be used in this context: The same filter that has been used to analyze real data can be applied to the simulated data. By comparing the estimated state of the simulated particles to the true state of the particles in the experimental data, it is possible to determine how well the simulation matches reality. Kalman filters are suited in the context of HEP because they support using combinations of (noisy) measurements from different detectors to provide a more accurate estimate of the particle's state. The advantage of the Kalman approach over the previous two approaches is that fake simulation will never be chosen as correct solution even if many malicious nodes are conspiring to publish identical results. A disadvantage to this approach is that anyone who has access to the real data of a given experiment might be able to efficiently fabricate simulation data that matches real-data without running the actual simulation. Another possible disadvantage here is that in rare cases, e.g. new physics might have been discovered in the real data that is used as reference, the simulations might be identified as being false due to the difference of Standard Model expectations (simulation data) and real data observed. In practice, such a scenario could be handled in the implementation of HEPchain by detecting unusual situations in which many simulations are uploaded by nodes but not a single correct one is among them. Then, root nodes might decide to fall back to one of the previously mentioned verification mechanisms or decide to run the simulation themselves.

\subsection{Properties of HEPchain}
1) Block sensitivity: Event generators like Pythia support setting a seed value that influences pseudo-random numbers required for simulation of probabilistic processes. If the seed value required for solving the next problem is (partly) determined by the hash of the previous block, then block sensitivity is given and neither is it likely that the same seed occurs multiple times nor is it possible to guess future problems which prevents pre-calculation of future results.\\\\
2) Adjustable problem hardness: Simulations of probabilistic processes are a trade-off between accuracy and computational effort. The problem hardness therefore can trivially be adjusted by lowering/raising energy cuts after which particles are dropped from the simulation. Note that this is just one of many viable techniques that can be used to influence the computational complexity of the simulations, one might as well adjust other parameters such as the amount of particle collisions or collision energies.\\\\
3) Parallelization: MC simulations by default are of so-called \textit{embarrassingly parallel} nature. As a result, the PoUW can be implemented so that it can make efficient use of multi-core CPUs and GPUs.\\\\
4) Fast Verification: Verification of results is the main challenge that needs to be overcome when transitioning from hash-based PoW to PoUW. In the context of a HEP PoUW, there are multiple reasons for this: For one, the resulting data from MC simulations (especially when the amount of simulated particles is set to a high number) is large compared to the fixed amount of bytes a cryptographic hash function outputs. This can lead to networking bottlenecks that might make traditional verification approaches infeasible, especially when considering that every node would first have to download all data that other nodes produce before being able to verify their correctness. A workaround here can be to increase the degree of centralization and promote an entity that profits from the useful work to a trusted root-authority that collects results from all miners and verifies the validity of their results. If spam attacks in which nodes upload false data become a problem, entering the blockchain can be coupled to the registration of real identities so that malicious nodes can be banned from participating in the blockchain. While it is true that increasing the amount of centralization in the context of blockchains that were invented with a decentralized nature in mind seems counter-intuitive, the reality is that in PoW like Bitcoin the three largest mining pools Foundry USA, AntPool and F2Pool as of January 2023 are estimated to control more than half of the total network hash rate. \cite{bitcoin_miningpools}\\
In addition to these aspects, a (potentially probabilistic) verification algorithm that given a proposed solution is able to identify the parameters used to generate this simulation data and to verify the correctness of calculations without repeating every calculation needs to be designed.

\section{Conclusion}
This publication has described the role of simulations in HEP and given an overview of distributed computing approaches commonly used. A new blockchain-based approach called HEPsim has been introduced and analyzed with regards to properties that traditional hash-based PoW has.\\It has been concluded that verification of results is the main challenge of designing an PoUW consensus algorithm for HEP. Different approaches that make use of validation-by-replication, in-time pre-calculation of results or comparisons to real data have been presented and their pros and cons have been traded off against each other.\\While the implementation of HEPchain is still in the works, its potential to support real-world experiments with valuable simulation or analysis data should not be underestimated. Its distinguishing factor compared to existing Volunteer Computing approaches (e.g. BOINC projects) is that it can easily be coupled with a cryptocurrency which provides might provide a financial incentive for miners to participate in the blockchain and support a real-world HEP experiment. It remains to be seen how efficiently verification of results can be implemented and how accurate validity predictions of given simulation data will be in practice.
\section{Future work}
\bibliographystyle{unsrt}
\bibliography{references}

\end{document}